\documentclass[12pt]{iopart}

\usepackage{txfonts}
\usepackage{graphicx}
\usepackage{subfigure}
\usepackage{color}

\usepackage{hyperref}
\hypersetup{colorlinks=true, linkcolor=blue, citecolor=blue, urlcolor=blue}

\newcommand{\mb}{\mathbf} 

\usepackage{cite}
\begin{document}

\title{MeV femtosecond electron pulses from direct-field acceleration in low density atomic gases}

\author{Charles Varin$^1$, Vincent Marceau$^2$, Pascal Hogan-Lamarre$^2$, Thomas Fennel$^3$, Michel Pich\'e$^2$, and Thomas Brabec$^1$}
\address{$ˆ1$ Center for Research in Photonics, University of Ottawa, Ottawa (ON) K1N 6N5, Canada}
\address{$ˆ2$ Centre d'Optique, Photonique et Laser, Universit\'e Laval, Qu\'ebec (QC) G1V 0A6,
Canada}
\address{$ˆ3$ Institut f\"{u}r Physik, Universit\"{a}t Rostock, 18051 Rostock, Germany}
\ead{cvarin@uottawa.ca}

\begin{abstract}
Using three-dimensional particle-in-cell simulations, we show that few-MeV electrons can be produced by tightly focusing few-cycle radially-polarized laser pulses in a low-density atomic gas. In particular, it is observed that for the few-TW laser power needed to reach relativistic electron energies, longitudinal attosecond microbunching occurs naturally, resulting in femtosecond structures with high-contrast attosecond density modulations. The three-dimensional particle-in-cell simulations show that in the relativistic regime the leading pulse of these attosecond substructures survives to propagation over extended distances, suggesting that it could be delivered to a distant target, with the help of a properly designed transport beamline.
\end{abstract}

\maketitle
\ioptwocol

\section{Introduction}
Optical electron acceleration is an important application of high-power lasers, in particular for the development of new electron and radiation sources for probing ultrafast structural dynamics in matter~\cite{carbone2012_cp,hada2013_epjst}. So far, research has mainly focused on the lower and upper limits of the energy spectrum:  subrelativistic energies for ultrafast electron diffraction (UED)~\cite{tokita09_apl,uhliga11_lpb,payeur12_apl,he13_njp,breuer2013_prl,marceau2013_prl,marceau2015_jpb} and ultrarelativistic energies for the development of compact radiation sources~\cite{schlenvoigt2007_natphys,fuchs2009_natphys,kneip2010_natphys,cipiccia2011_natphys}. In comparison, little attention has been paid to optical electron acceleration in the few-MeV range. 

MeV electrons are of interest for relativistic time-resolved electron diffraction~\cite{hastings2006_apl,MurookaAPL2011}, as well as for controlled injection in staged laser wakefield accelerators~\cite{malka2012_pop,hooker2013_nature} and dielectric acceleration structures~\cite{peralta2013_nature,breuer2013_prl,England2014} that in turn can drive table-top synchrotron radiation sources~\cite{schlenvoigt2007_natphys,fuchs2009_natphys,kneip2010_natphys,cipiccia2011_natphys}. In particular for diffraction with ultrashort electron pulses, the suppression of space-charge effects at relativistic energies allows packing more electrons per bunch while maintaining ultrashort duration at a distant target. This enables single-shot studies of ultrafast structural processes that are out of reach of conventional (non-relativistic) UED~\cite{sciaini2011_rpp}.

There actually exist ultrafast sources of MeV electrons. For example, radio frequency (rf) photoguns can produce few-MeV pulses with 100-fs duration~\cite{MurookaAPL2011,musumeci2010_apl,hada2013_epjst} and there are proposals to obtain even shorter durations ($\sim$10 fs) with an rf compressor~\cite{han11_prstab}. However, using a laser-based acceleration mechanism has the following potential advantages: (i) the short wavelength of the accelerating field may lead to electron bunches with initial durations of the order of 10 fs or less; (ii) the intrinsic synchronization between the laser and electron pulses is essentially jitter free; (iii) compared with nanoclusters and thin film targets, optimization using a gas medium does not require sub-wavelength alignment of the target position; (iv) using a gas medium, the target is self-regenerating and can thus be used for experiments at high repetition rates. Breakthrough laser-driven plasma acceleration experiments demonstrated the production of fs electron pulses in the sub-100-MeV range~\cite{schmid2009_prl,Lundh2011} and a method to obtain 15-fs pulses in the 5-10 MeV range was proposed~\cite{beaurepaire14_njp}. 

In this article, we show how ultrashort MeV electron pulses can be produced by tightly focusing few-TW radially-polarized femtosecond laser pulses (RPFLPs) in a low-density gas. 
We investigate the laser-gas interaction and electron acceleration in a configuration similar to the experiment reported by Payeur~\emph{et al.}~\cite{payeur12_apl} with three-dimensional particle-in-cell (3DPIC) simulations. In particular, we observed attosecond microbunching within a longer femtosecond structure for a laser power in the few-TW range. This is about two orders of magnitude lower than previously demonstrated for direct-field electron acceleration~\cite{varin2006_pre,karmakar2007_lpb}. We show that some of these microbunches survive to propagation in space, suggesting that the proposed method could enable the delivery of sub-femtosecond electron pulses to a distant target with the current laser and beamline technology.

This paper is divided as follows. First in Sec.~\ref{sec:method}, we present our 3DPIC modelling approach to relativistic RPFLP-gas interaction. The analysis includes an exact analytical solution to Maxwell's equations for the spatiotemporal distribution of tightly-focused RPFLPs and the complete ionization and plasma dynamics starting from a neutral atomic gas. In Sec.~\ref{sec:conditions}, we explain why MeV electron energies and attosecond bunching are expected in the few-TW regime under tight focusing conditions. Analytical predictions are confirmed with 3DPIC simulations. In Sec.~\ref{sec:propagation}, we analyze the propagation of the electron pulses and explain how the electron distributions could be filtered to isolate the leading substructure. In Sec.~\ref{sec:quality}, we discuss the quality of that substructure and its distribution in phase-space to show that it is appropriate for being temporally compressed and transversely focused.
Finally, we draw conclusions in Sec.~\ref{sec:conclusion}.

\section{Relativistic laser-gas interaction model\label{sec:method}}
To simulate the complete ultrafast and ultra-intense RPFLP-gas interaction, we performed 3DPIC simulations with full atomic ionization and charge dynamics. In particular, we used a customized version of the EPOCH PIC code~\cite{brady11_ppcf}, where the electromagnetic field of the RPFLP is introduced using the FDTD scattered-field formulation~\cite{taflove05_book}. This approach allows us to provide the most accurate description of tightly-focused RPFLPs \emph{via} the analytical vacuum-field solution.

For the vacuum field representing the incident RPFLP, we use an exact closed-form spatiotemporal formulation given by the following expressions in cylindrical coordinates, with $\mathbf{r}=(r,\theta,z)$~\cite{april10_intech,marceau12_optlett}:
\begin{eqnarray}
E_r (\mb{r},t) =  \frac{3 E_0 \sin 2\tilde{\Theta}}{2\tilde{R}} \bigg( \frac{G_-^{(0)}}{\tilde{R}^2} + \frac{G_+^{(1)}}{c\tilde{R}} + \frac{G_-^{(2)}}{3c^2}\bigg) \ ,  \label{eq:npTM01Er}\\
E_z (\mb{r},t) =   \frac{E_0}{\tilde{R}} \bigg[  \frac{(3\cos^2\tilde{\Theta}-1)}{\tilde{R}}  \bigg( \frac{G_-^{(0)}}{\tilde{R}} + \frac{G_+^{(1)}}{c} \bigg) \nonumber \\ 
\qquad \qquad \qquad - \frac{\sin^2\tilde{\Theta}}{c^2} G_-^{(2)} \bigg]  \ , \label{eq:npTM01Ez} \\ 
B_\phi (\mb{r},t) =  \frac{E_0 \sin \tilde{\Theta}}{c \tilde{R}} \bigg( \frac{G_-^{(1)}}{c\tilde{R}} + \frac{G_+^{(2)}}{c^2}\bigg)  \ . \label{eq:npTM01Bphi}
\end{eqnarray}
The physical fields $\mathbf{E}$ and $\mathbf{B}$ are obtained by taking the real part of the corresponding expressions. $E_0$ is the amplitude of the transverse electric field component and $c$ is the speed of light \emph{in vacuo}. The complex variables are defined as follows: $\tilde{R}=[r^2 + (z+ja)^2]^{1/2}$, $\cos \tilde{\Theta} = (z+ja)/\tilde{R} $, and $G^{(n)}_\pm = \partial^n_t [f(\tilde{t}_-)\pm f(\tilde{t}_+)]$ with $\tilde{t}_\pm = t \pm \tilde{R}/c + ja/c$. The confocal parameter $a$ is associated with the Rayleigh range $z_R$ by $k_0 z_R = \sqrt{1+(k_0 a)^2} - 1$~\cite{rodriguez-morales04_optlett}, where $k_0 = 2\pi/\lambda_0$, $\lambda_0$ being the central wavelength of the laser. 

A Poisson-like pulse spectrum~\cite{caron99_jmodoptic} was assumed, with the corresponding temporal form $f(t) = e^{-j\phi_0}\left(1-j\omega_0 t/s\right)^{-(s+1)}$,
where $\phi_0$ is the constant pulse phase and $\omega_0=ck_0$ the central angular frequency of the pulse spectrum. The parameter $s$ is a positive constant related to the pulse duration. In the limit where $s\gg 1$, the usual carrier-envelope temporal function  $f(t)\simeq e^{-t^2/T^2} e^{j(\omega_0 t- \phi_0)}$ is recovered, with $T \simeq \sqrt{2s}/\omega_0$. From an experimental point of view, the fields given by Eqs.~(\ref{eq:npTM01Er})--(\ref{eq:npTM01Bphi}) can be produced by focusing a collimated radially-polarized input beam with a parabolic mirror of large numerical aperture~\cite{april10_optexpress,payeur12_apl}. 

\begin{figure}[!t]
\centering 
\includegraphics[width=0.9\columnwidth]{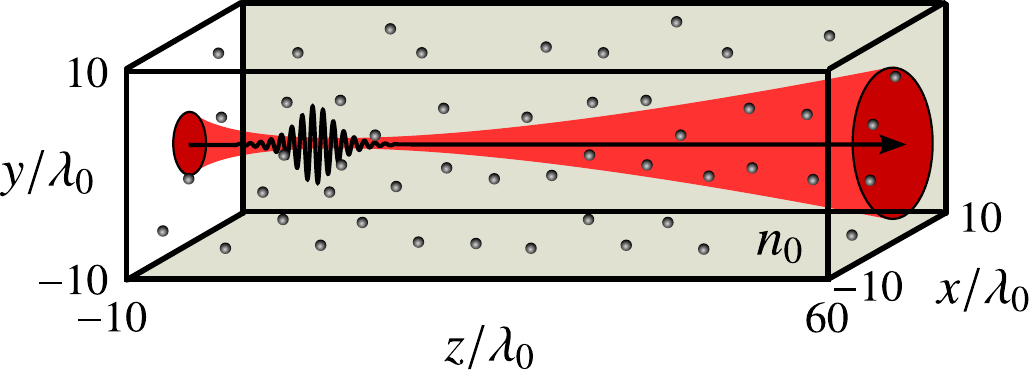}
\caption{Schematic representation of the laser-gas interaction volume analyzed with 3D particle-in-cell simulations (3DPIC).\label{fig:scheme}}
\end{figure}

Typical numerical simulations are done with a cartesian mesh of size $600 \times 600 \times 2100$ with $N_\lambda=30$ grid points per wavelength, spanning the region $x,y \in [-10\lambda_0, 10\lambda_0]$ and $z \in [-10\lambda_0, 60\lambda_0]$. We have observed that increasing the grid resolution to $N_\lambda=40$ did not change the results significantly. At $t=0$, the laser pulse is outside the simulation domain and the gas is represented by neutral atoms randomly distributed in the interaction zone and modeled by some $10^8$ particle markers. A schematic illustration of the simulation setup is given in Fig.~\ref{fig:scheme}.

Throughout the simulations, multiphoton, tunnel, and barrier-suppression ionization were taken into account, as well as the motion of the ions and its effect on the electron trajectories (implementation details are found in~\cite{lawrence_phdthesis}). Due to the low density of the ionized medium, collisions and electron impact ionization were neglected. 

We recall that in analogy to the standard normalized transverse vector potential parameter $a_0 = e|E|/m_ec\omega_0$~\cite{hartemann01}, when considering an RPFLP it is useful to introduce a normalized longitudinal parameter $a_z = e|E_z|_{peak}/m_ec\omega_0$~\cite{varin13_applsci}, where $|E_z|_{peak}$ is the amplitude of the longitudinal electric field component given at Eq.~(\ref{eq:npTM01Ez}). The relativistic regime associated with $a_z\gg 1$ is characterized by sub-cycle electron microbunching, explained below (see also~\cite{varin2006_pre,varin13_applsci}).

In direct-field acceleration, the direction of the initial longitudinal push to the electrons depends on the sign of the local electric field. Assuming an infinite homogeneous electron gas, half of the free electrons will start moving along the laser pulse propagation vector, whereas the other half will move in the opposite direction. Effectively, the electrons in each field cycle will converge locally to the optimal acceleration phase position. In the weak field limit ($a_z \ll 1$), the expected periodic modulation of the longitudinal electron density is imperceptible. Nevertheless, in the ultra-relativistic regime ($a_z \gg 1$), the modulation is strong and seen as a series of microbunches, separated by $\lambda_0$. When the microbunches are locked to the phase of the laser pulse and accelerated to relativistic energies, they are further compressed by the local longitudinal electric field, leading to microbunches of attosecond duration.

\section{Direct-field electron acceleration and attosecond bunching with RPFLPs tightly focused in a low-density gas\label{sec:conditions}}

In previous publications~\cite{marceau2013_prl,marceau2015_jpb}, it was shown that femtosecond few-hundred-keV electron pulses with percent-level energy spread can be produced by tightly focusing few-mJ RPFLPs in an hydrogen gas whose density is in the $10^{22}\,\mathrm{m}^{-3}$ range. In particular, it was demonstrated that the laser pulse parameters and gas density can be optimized to cover the 100 -- 300~keV energy window that characterizes ultrafast electron diffraction imaging experiments~\cite{sciaini2011_rpp}. 

At a dominant wavelength of $\lambda_0 = 800$~nm, typical pulse parameters were $k_0a = 20$, $s = 70$, and $P=300$~GW, corresponding to a few-mJ 8-fs full width at half maximum (FWHM) pulse focused to a $2\lambda_0$ spot diameter. These parameters, defined by an extensive set of optimization calculations~\cite{marceau2015_jpb}, were found to offer a good compromise between transverse and longitudinal effects, with a positive impact on electron energy, bunch charge, divergence, and energy spread. 

In the regime explored in~\cite{marceau2013_prl,marceau2015_jpb}, the normalized longitudinal parameter $a_z$ is slightly larger than unity. In such conditions, the longitudinal electric field is too weak to induce a strong longitudinal compression and attosecond bunching. Instead, the ultrashort electron pulse duration is due to the acceleration of a thin disk of electrons located in a very restricted region of the infinite gas target~\cite{marceau2013_prl}. Electrons outside this thin disk region are either deflected away from the optical axis or gain little energy from the laser field. This effectively results in electron pulses with initial sub-femtosecond duration. However, the duration of these sub-relativistic pulses increases rapidly as they propagate due to electrostatic repulsion and the finite initial energy spread.

\begin{figure}[!t]
\centering 
\includegraphics[width=\columnwidth]{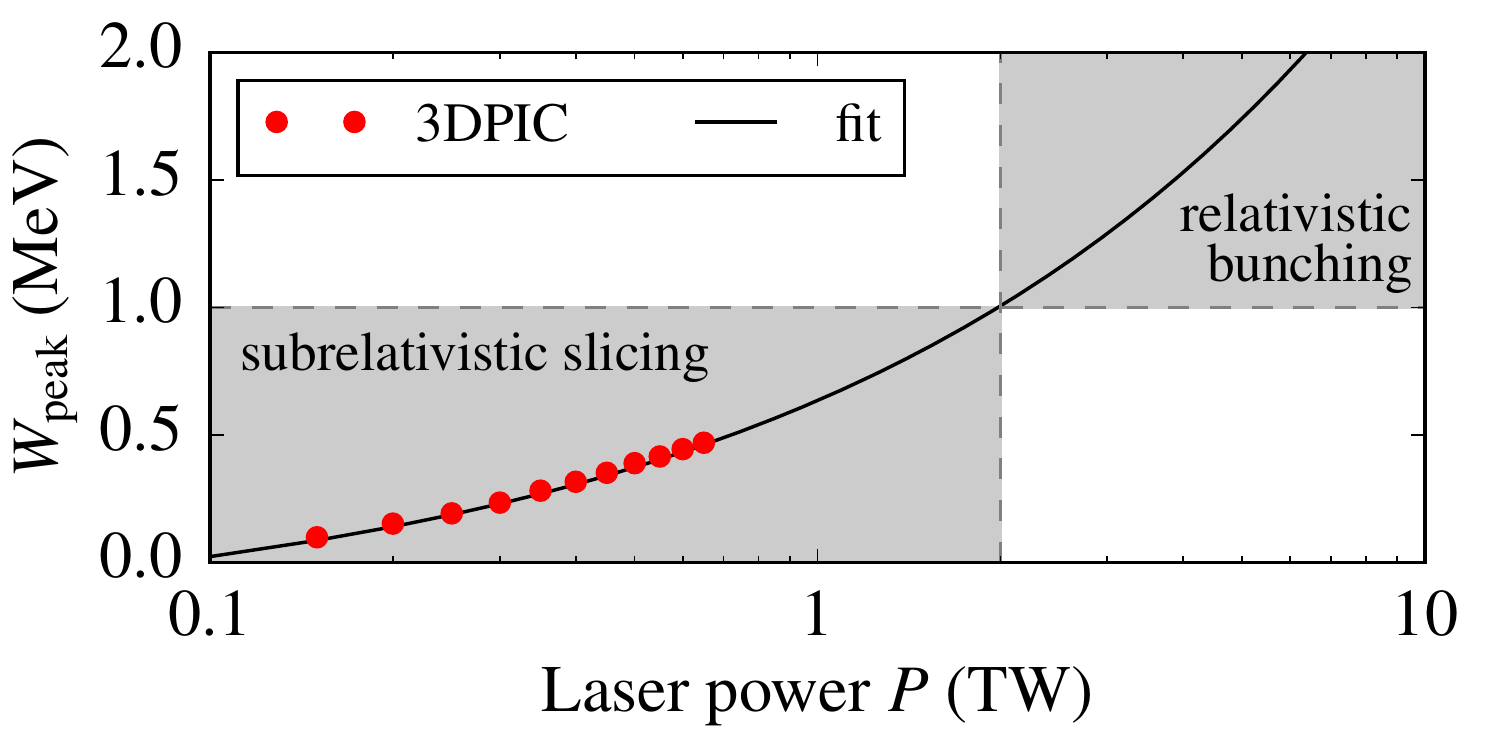}
\caption{Analysis of electron acceleration in tightly-focused RPFLP as a function of the laser power. Red dots (\textcolor{red}{$\bullet$}) are 3DPIC data points from Fig. 5(a) of \cite{marceau2015_jpb}, while the solid black line (--) is a fit of the form $a + b\sqrt{P}$ (see also \cite{fortin10_jpb}).  We identified in gray two different ultrashort electron pulse regimes: the subrelativistic slicing regime, where only a thin disk of electron is accelerated (see, e.g., \cite{marceau2013_prl}), and the relativistic bunching regime, where electrons are axially compressed into attosecond pulses (see, e.g., \cite{varin13_applsci}).\label{fig:scaling}}
\end{figure}

\begin{figure*}[!t]
\centering 
\includegraphics[width=\textwidth]{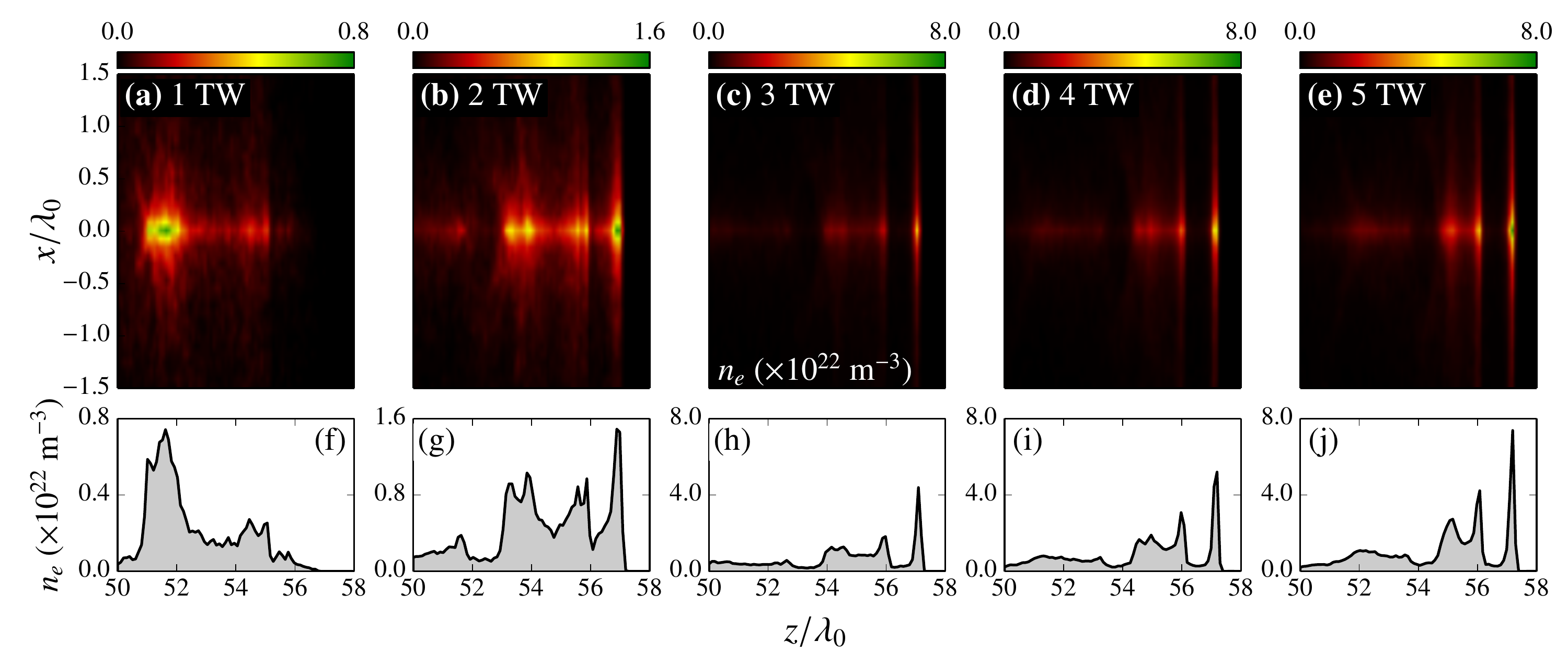}
\caption{3DPIC analysis of the transition to the relativistic regime of direct-field electron acceleration by 8-fs (FWHM) RPFLPs tightly focused in a low density helium gas ($k_0a = 20$, $s = 70$, $\phi_0 = \pi$, and $n_{\mathrm{He}}=10^{22}\,\mathrm{m}^{-3}$). (a)-(e) are 2D cuts of the 3D electron density structures away from the RPFLP focus along the $z$ coordinate for different values of the laser power. (f)-(j) are 1D cuts at $x=0$ of the corresponding top graphs. Looking at (a) to (c) and (f) to (h), one sees clearly the transition from the subrelativistic slicing regime to the 
relativistic attosecond bunching regime, where the density distribution exhibits a sharp modulation at the leading edge due to sub-cycle compression of the accelerated electrons. Below the relativistic compression threshold in (a)-(b), the electron density has a peak value less than the initial bound-electron density ($n_e=2n_{\mathrm{He}}$). However in (c)-(e) above threshold, the compressed electron densities have peak values greater than $2n_{\mathrm{He}}$. Still in (c)-(e), it is observed that the peak density increases with the laser power, without important changes in the macrobunches structure. Ultimately in (e), the peak charge density exceeds that of (a) by an order of magnitude. Total charge in the 3D macrobunches was found to be approximately (a) 1~fC, (b) 2~fC, (c) 3~fC, (d) 4~fC, and (e) 5~fC. Charge in the leading bunch is about (b) 0.35~fC, (c) 0.67~fC, (d) 0.85~fC, and (e) 1.05~fC.\label{fig:transition}}
\end{figure*}

To predict the conditions to reach the relativistic regime---where longitudinal compression would become important---we fitted results from~\cite{marceau2015_jpb} for the electron energy as a function of the power $P$ of the incident laser pulse. We used a $\sqrt{P}$ scaling law that characterizes direct-field acceleration (see, e.g.,~\cite{fortin10_jpb}). Results are shown in Fig.~\ref{fig:scaling}. Conclusions are: (1) average MeV energies should be reached when $P \gtrsim 2\,\mathrm{TW}$ and (2) for strong focusing conditions ($z_R \sim \lambda_0$), in the few-TW power range, the longitudinal electric field has a super-relativistic strength ($a_z\gg 1$). According to the 1D longitudinal direct-field acceleration model~\cite{varin13_applsci}, longitudinal attosecond bunching should then occur naturally.

To validate our predictions, we simulated electron acceleration in helium with the 3DPIC method discussed in Sec.~\ref{sec:method}. For direct comparison with the results obtained in the subrelativistic electron regime~\cite{marceau2013_prl,marceau2015_jpb}, we used the following pulse parameters: $k_0a = 20$, $s = 70$, and $\phi_0=\pi$, with a laser power in the 1-5 TW range. 

Results from the 3DPIC analysis are shown in Fig.~\ref{fig:transition}. They reveal that a high-contrast attosecond modulation of a longer femtosecond electron pulse structure is effectively produced when the laser power is in the few TW range. This corresponds to sub-100-mJ laser pulse energy. We recall that previous demonstrations assumed Joules (few-cycle pulses, 100~TW) or more to produce similar attosecond bunching~\cite{varin2006_pre,karmakar2007_lpb}. In these studies, nanoclusters were considered to restrict the initial volume occupied by the free electrons. Here, the attosecond electron microbunches are produced in an infinite gas target. This configuration is much simpler, as it does not require particular target alignment. 

\begin{figure}[!t]
\centering 
\includegraphics[width=\columnwidth]{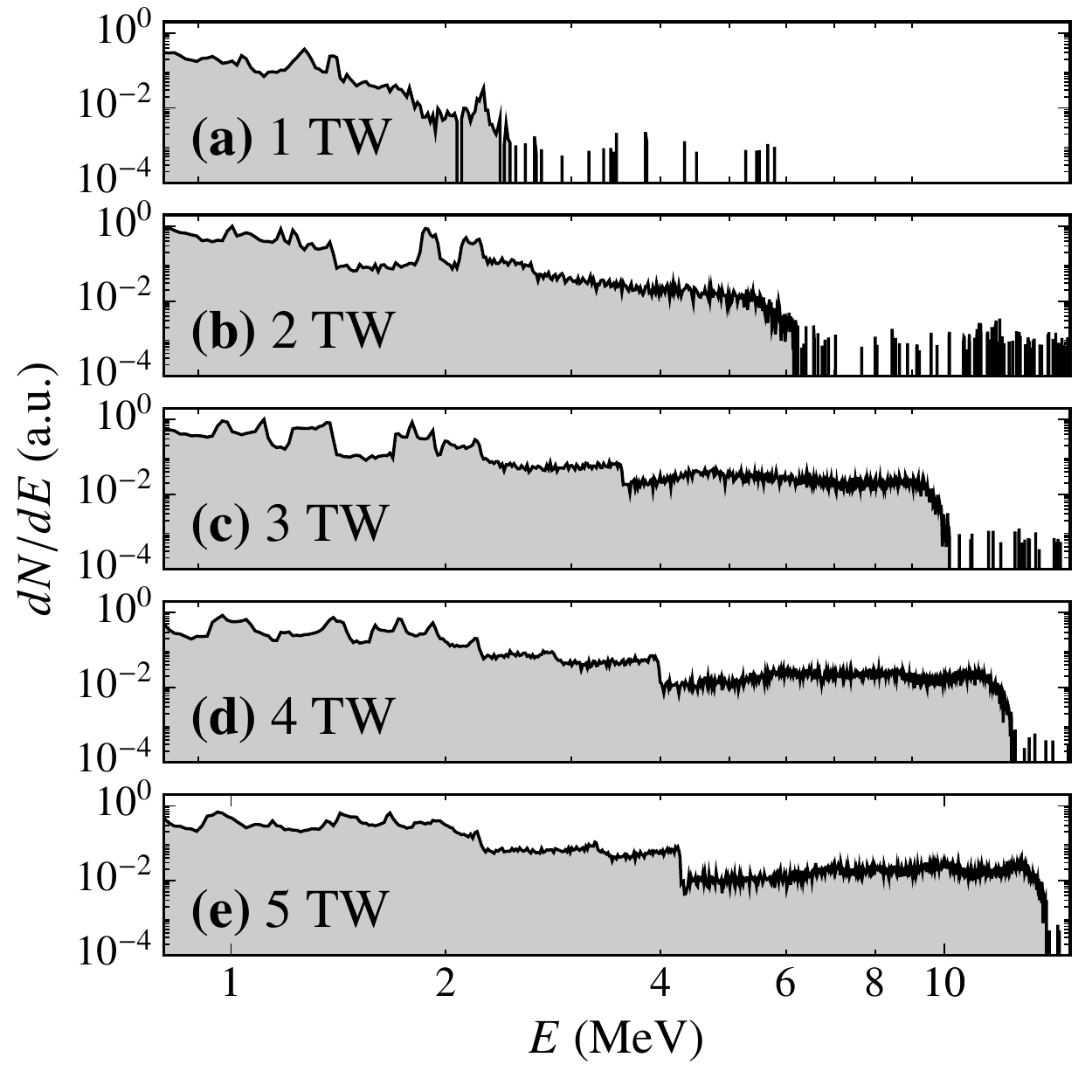}
\caption{The broad energy spectra associated with the femtosecond electron structures shown in Fig.~\ref{fig:transition}.\label{fig:spectra}}
\end{figure}

From the spectra given in Fig.~\ref{fig:spectra}, it is evident that the femtosecond electron structures shown in Fig.~\ref{fig:transition} are composed of electrons with a broad distribution of energy. They are thus unlikely to remain together over an extended period of time. However, the spectral analysis of particular substructures (not shown) suggests that these complex femtosecond pulses are effectively composed of individual subpulses. We analyze this aspect in the following section by looking at the evolution of the macrobunches over time, e.g., during propagation to a distant target.

\section{Analysis of the propagation to a distant target\label{sec:propagation}}

Electrostatic repulsion and electron momentum dispersion cause the spatio-temporal broadening of an electron pulse in time~(see, e.g.,~\cite{Michalik2009}). For ultrashort pulses, the effect is predominantly longitudinal, causing the increase of the duration as the bunch travels in space~\cite{Siwick2002}. In case of a broad spectrum, like that shown in Fig.~\ref{fig:spectra}, an electron pulse might literally break up before it can be delivered to a distant target, e.g., where it would interact with a sample to produce a useful diffraction pattern.

We analyzed the propagation of the electron pulse structures shown in Fig.~\ref{fig:transition} with 3DPIC by using the moving window technique, were the numerical mesh is translated at the speed of light to follow the electron pulses as they move. The maximum physical time and distance that can be simulated essentially depend upon the available computational resources. With the parameters given in Sec.~\ref{sec:method}, but where $x,y \in [-5\lambda_0, 5\lambda_0]$ to reduce the computational load, simulating a 700-fs propagation takes about 40 hours on a supercomputer using some 128 processes. It corresponds to a distance of 0.2~mm away from the RPFLP focus, comparable to the longitudinal distances reached in similar PIC studies~\cite{beaurepaire14_njp}.

\begin{figure*}[!t]
\centering 
\includegraphics[width=0.8\textwidth]{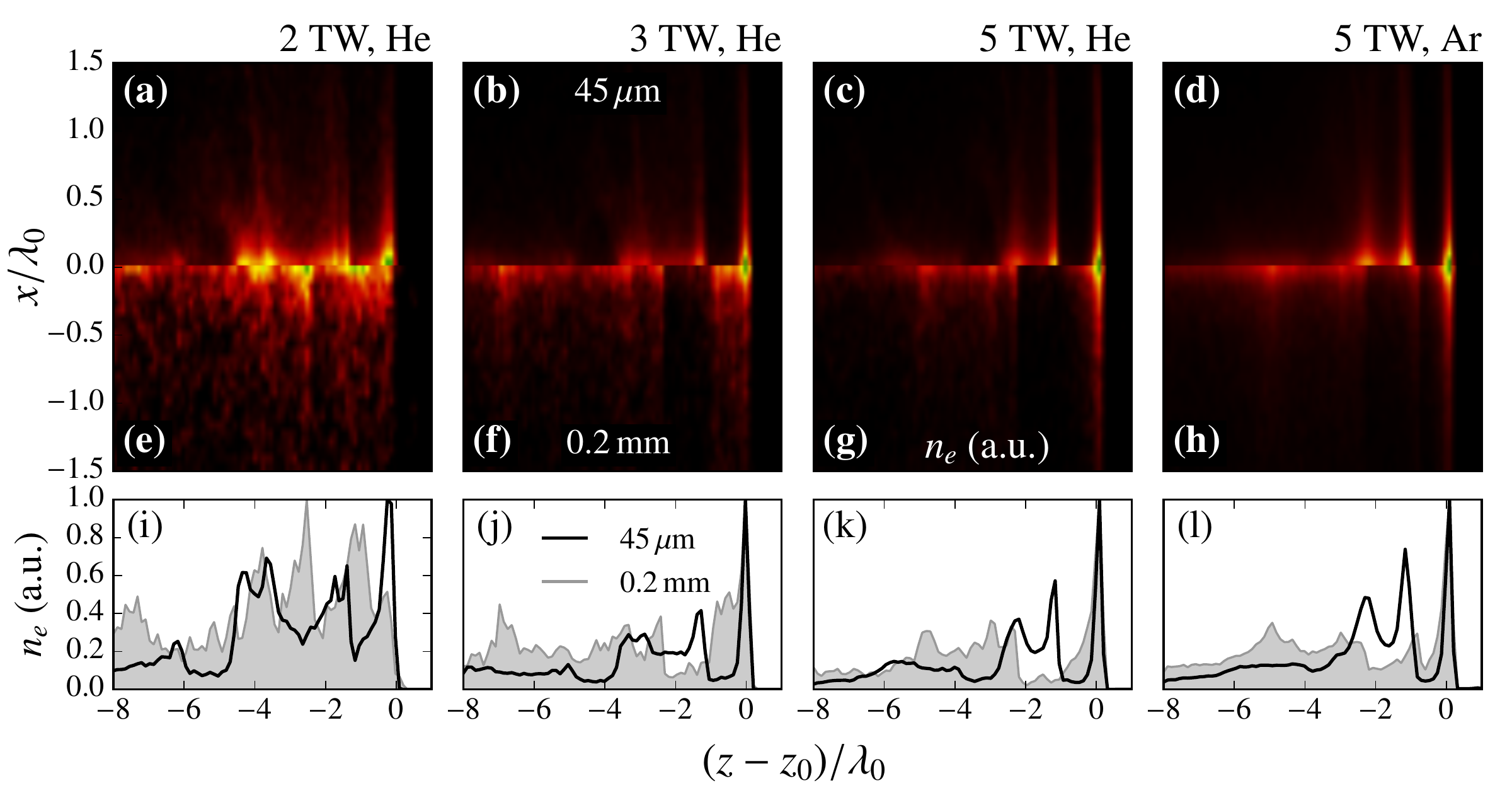}
\caption{3DPIC analysis of the evolution of the relativistic electron macrobunches produced by a few-TW 8-fs (FWHM) RPFLP tightly focused in a low density gas ($k_0a = 20$, $s = 70$, $\phi_0=\pi$, and $n=10^{22}\,\mathrm{m}^{-3}$). Color plots show 2D cuts of the 3D electron density distributions $n_e$ away from the optical focus. (a)-(d) $z_0 = 45\,\mu\mathrm{m}$ and (e)-(h) $z_0 = 0.2\,\mathrm{mm}$. (i)-(l) are 1D cuts at $x=0$ of the same distributions: $45\,\mu\mathrm{m}$ as a solid black line and $0.2\,\mathrm{mm}$ as a gray area. All distributions were normalized for direct comparison. For $P = 2\,$TW [(a), (e), and (i)], the structure spreads out and the initial longitudinal modulation disappears. For $P = 3\,$TW [(b), (f), and (j)], the leading attosecond pulse is still apparent after propagation, but with a much broader base. For $P = 5\,$TW [(c), (g), and (k)], the trailing structure of slow electrons spreads out, but broadening of the leading pulse base is limited. For $P = 5\,$TW, comparison between simulations with helium [(c), (g), and (k)] and argon [(d), (h), and (l)] shows that the formation of the leading bunch does not depend on a specific gas element. The initial bunch charge in (d) was about 37~fC, 7 times higher than that of the bunch shown in Fig.~\ref{fig:transition}(e).\label{fig:propagation}}
\end{figure*}

The results given in Fig.~\ref{fig:propagation} show that at the relativistic threshold (2~TW), the initial density modulations are lost during propagation. However, above relativistic threshold ($>$ 2~TW) the leading substructure survives. For the electrons accelerated by the 5-TW RPFLP, it even maintained a sub-femtosecond duration. In comparison, a subrelativistic bunch with a comparable initial duration and charge would have a duration in the tens-of-fs range (see, e.g.,~\cite{marceau2015_jpb}). 

The macrobunches produced in the longitudinal compression regime share these common features: a fast and sharp leading structure followed by a slower trailing complex. We have observed that by applying a high-pass energy filter, the leading pulse structure is isolated. Conceptually, such an energy filter is implemented by deflecting the electrons with a magnetic field and putting an obstacle in the path of the electrons with energies that need to be removed. This approach was demonstrated in the femtosecond regime by filtering the electron energy directly into a magnetic compression device~\cite{tokita10_prl}. In principle, the same technique can be used for attosecond electron pulses (see, e.g., \cite{Hansen2012}), although the efficiency of the approach in the multi-electron regime still needs to be demonstrated.

\begin{figure}[!t]
\centering 
\includegraphics[width=\columnwidth]{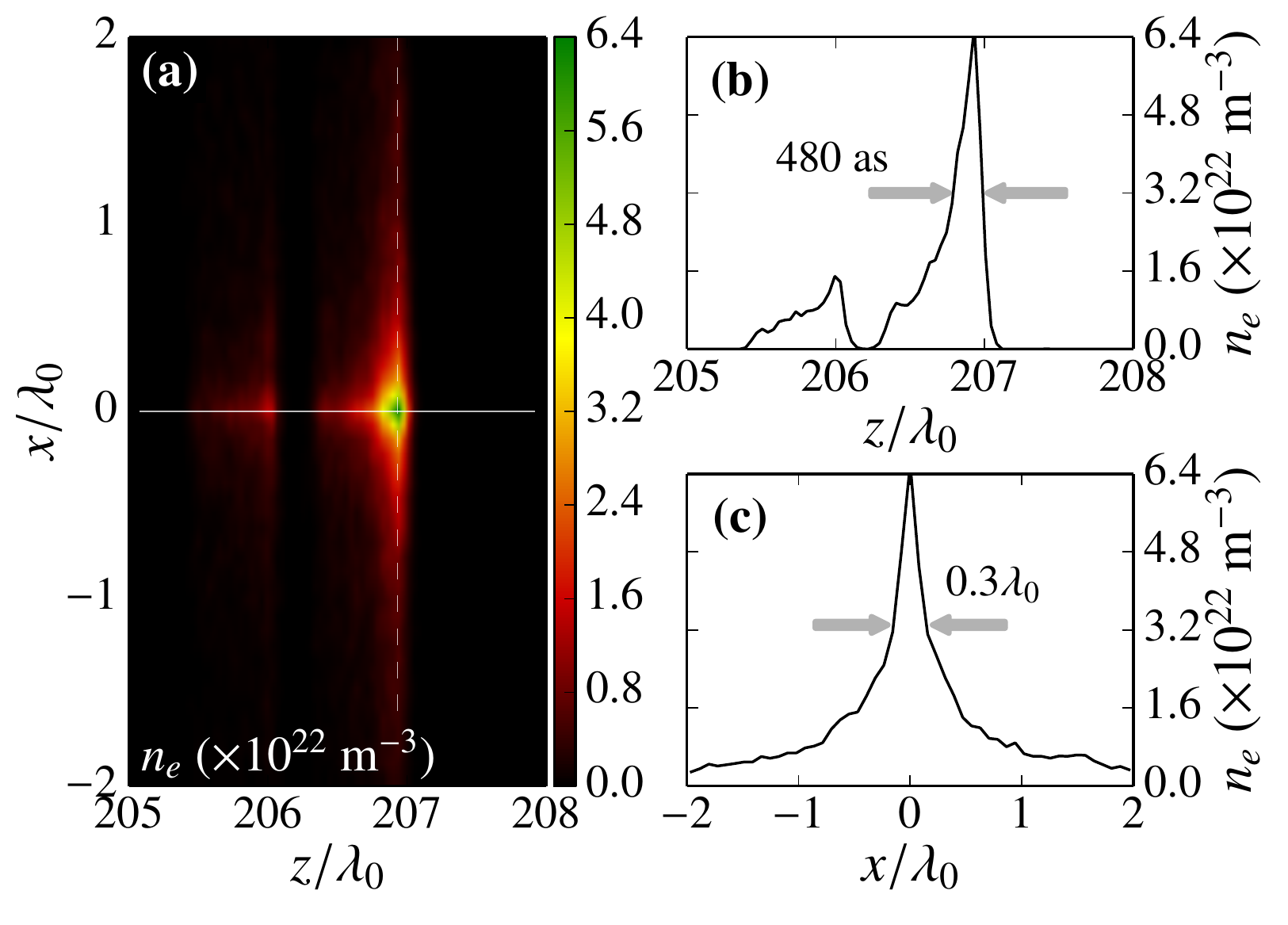}
\caption{Spatio-temporal distribution of the macrobunch of Fig.~\ref{fig:propagation}(h) after electrons with energy less than 5~MeV where filtered out. A 2D cut of the 3D electron density structure as well as 1D longitudinal and transverse cuts are shown in (a), (b), and (c), respectively. In (b) it is shown that even if the bunch is effectively 0.2~mm away from the optical focus, the FWHM duration is still in the attosecond range. Calculated bunch charge is 1.5~fC and average energy is 10~MeV.\label{fig:filtered_bunch}}
\end{figure}

To simulate the impact of an energy filter, we applied a numerical energy cutoff to the bunch structure shown in Fig.~\ref{fig:propagation}(h), by rejecting all the electrons with an energy less than 5~MeV. The resulting bunch, shown in Fig.~\ref{fig:filtered_bunch}, has a 480-as FWHM duration, 0.2~mm away from the optical focus. This unexpectedly short duration cannot be explained by relativistic effects alone.

Using the average energy and energy spread extracted from the 3D electron bunch of Fig.~\ref{fig:filtered_bunch}, respectively $E=10.2~$MeV and $dE=4.3~$MeV, we calculated the pulse duration predicted by the relativistic single-electron model presented in appendix. With $L=155~\mu\mathrm{m}$ [the distance between the two frames shown in Figs.~\ref{fig:propagation}(d) and~\ref{fig:propagation}(h)], it gives a pulse duration of approximately 1~fs, which overestimates the 3DPIC prediction significantly. We conclude that the space-charge fields from the highly charged trailing complex, which are neglected in the single-electron model, play a beneficial role in the current context by constraining the expansion of the leading attosecond substructure at the trailing edge.

Simulating macrobunch propagation with 3DPIC up to 10~cm, a typical distance in UED experiments~\cite{sciaini2011_rpp}, is impractical ($\sim$100 days with 1000 computer processes). However, the relativistic single-electron model for the bunch in Fig.~\ref{fig:filtered_bunch} gives an upper limit estimation of 350~fs. In comparison, a few-hundreds-keV pulse with a percent-level energy spread would be one to two orders of magnitude longer (see Eq.~(\ref{eq:spread_nr}) and, e.g., Fig.~13 in~\cite{marceau2015_jpb}). Still, the phase-space analysis we present next suggests that the energy distribution in the bunch is appropriate for temporal recompression, with the potential of sub-fs duration at realistic target distances.

\section{Analysis of the attosecond pulse quality\label{sec:quality}}

\begin{figure}[!t]
\centering 
\includegraphics[width=\columnwidth]{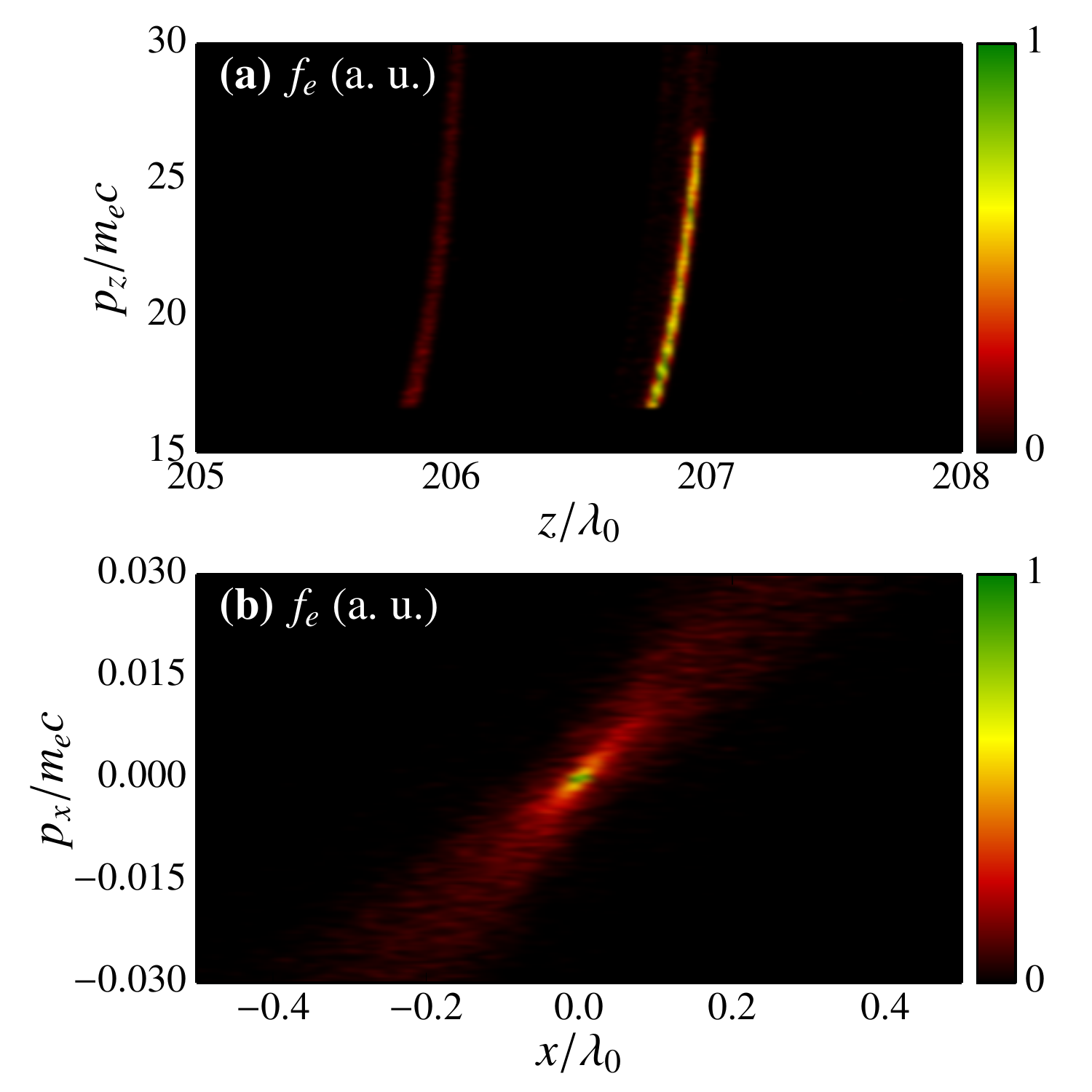}
\caption{Normalized (a) longitudinal and (b) transverse phase space distribution associated with the filtered electron pulse shown in Fig.~\ref{fig:filtered_bunch}. Both the longitudinal and transverse distributions exhibit a nearly linear chirp that is appropriate for pulse compression and focusing.\label{fig:filtered_phase_space}}
\end{figure}

We have seen in the previous section that the leading attosecond pulse shown in Fig.~\ref{fig:filtered_bunch} has a large relative energy spread ($dE/E\simeq 0.42$). However, a comparison between Eqs.~(\ref{eq:spread}) and~(\ref{eq:spread_nr}) indicates that the temporal spreading associated with this relativistic pulse is comparable to that of a 100-keV pulse with $dK/K\approx 10^{-3}$, which is very difficult to achieve ($K$ stands for the kinetic energy). Close inspection of the longitudinal phase-space distribution given in Fig.~\ref{fig:filtered_phase_space}(a) moreover shows that the energy is distributed almost linearly in the bunch, which is ideal for recompression. The distribution actually suggests that the duration of the leading structure after recompression could even be shorter than 480~as.

The thermal transverse emittance associated with the transverse phase-space distribution shown in Fig.~\ref{fig:filtered_phase_space}(b) is $\epsilon_{n,x}=1.5\times 10^{-2}\,\textrm{mm mrad}$, similar to known values for radio frequency photoinjectors (typically $\epsilon_{n,x} < 5\times 10^{-2}\,\textrm{mm mrad}$)~\cite{Li2012PRSTAB}. However, due to the large angular spread imprinted on the electron distribution by the curved phase front of the laser pulse, the transverse coherence length is small $L_T = \hbar/\sqrt{\langle p_x^2 \rangle} \sim 10^{-2}\,\mathrm{nm}$). For a demonstration UED experiment, the electron pulse would then need to be focused to have a coherence length that is sufficiently long to form useful images; typically $L_T > 1\,\mathrm{nm}$ is needed~\cite{sciaini2011_rpp}. We emphasize that focusing is necessary to keep a maximum of the charges in the bunch. But this operation is likely to increase the bunch duration, as the net focusing effect of the current magnetic lenses used in UED beamlines depends on the electron energy. With focusing in mind, it is probably safer to say that the final electron pulse duration would be in the few-fs range.

\section{Conclusion}\label{sec:conclusion}
Using three-dimensional particle-in-cell simulations, we have shown that relativistic electron pulses can be produced by tightly focusing few-cycle radially-polarized laser pulses in a low density atomic gas. In particular, we observed that for a laser power in the few-TW range, longitudinal attosecond microbunching occurs naturally, resulting in femtosecond structures with high-contrast attosecond density modulations. 

It was demonstrated that in certain conditions the leading attosecond substructure survives to propagation over millimeter-scale distances. Combined with a high-pass energy filter, focusing element, and dispersion compensator, the proposed method could allow relativistic electron diffraction with a temporal resolution in the few-fs range, potentially sub-fs.

Optimization of the laser pulse and gas parameters is currently being studied to improve the final bunch quality and total charge. Preferably, future work should include realistic modeling of the beamline to include the effects associated with the energy filter and focusing element. We emphasize that increasing the bunch charge is currently the most critical issue toward real application of the method. Stability of the bunch properties with respect to shot-to-shot fluctuations of the pulse energy and carrier-envelope phase should also be addressed.

\ack
This research was supported by the Natural Sciences and Engineering Research Council of Canada (NSERC). Computations were done on the supercomputer Briar\'ee from Universit\'e de Montr\'eal, managed by Calcul Qu\'ebec and Compute Canada. The operation of this supercomputer is funded by the Canada Foundation for Innovation (CFI), NanoQu\'ebec, RMGA and the Fonds de recherche du Qu\'ebec - Nature et technologies (FRQ-NT). The authors gratefully acknowledge the EPOCH development team. The EPOCH code used in this research was developed under UK Engineering and Physics Sciences Research Council grants EP/G054940/1, EP/G055165/1 and EP/G056803/1. 

The authors thank the reviewers for their insightful comments and suggestions as well as Fran\c{c}ois L\'egar\'e, Jean-Claude Kieffer, and Martin Centurion for stimulating discussions.

\appendix
\section*{Appendix}\label{appendix}
\setcounter{section}{1}
An electron with velocity $v$ at $z = 0$ will arrive at a remote position $z=L$ at time $t = L/v$. Using the usual relativistic formulas $v = c^2p/E$ and $E = \sqrt{c^2p^2 + m^2c^4}$, this time can be written as $t = (L/c)\times(E/\sqrt{E^2 - m^2c^4})$. Then, the arrival time delay associated with a collection of non-interacting electrons whose energies are within a certain energy range $dE$ around $E$ is, in the relativistic limit ($E \gg mc^2$):
\begin{equation}\label{eq:spread}
d t = \frac{L}{c}\left[\frac{m^2c^4}{\left(E^2-m^2c^4\right)^{3/2}}\right]dE \simeq \frac{L}{c}\frac{m^2c^4}{E^3}dE.
\end{equation}
An estimation of the pulse duration at the target due to the finite energy spread is then $\Delta t + dt$, where $\Delta t$ is the initial pulse duration. It is straightforward to show that the non-relativistic form of Eq.~(\ref{eq:spread}) ($K \ll mc^2$) is:
\begin{equation}\label{eq:spread_nr}
d t = \sqrt{m}\frac{L}{(2K)^{3/2}}dK,
\end{equation}
where $K$ stands for the kinetic energy.


\providecommand{\newblock}{}

\end{document}